\documentclass[fleqn,twoside]{article}
\usepackage[headings]{espcrc2}
\usepackage{pstricks}
\readRCS
$Id: espcrc2.tex,v 1.2 2004/02/24 11:22:11 spepping Exp $
\usepackage{graphicx}

\usepackage[figuresright]{rotating}
\newcommand{\ra}{\rightarrow}

\newcommand{\Cascade}{\ifmmode  \Xi%
                 \else%
                            \mbox{$\Xi$}%
                 \fi%
                 }%

\newcommand{\om}{\ifmmode  \Omega%
                 \else%
                            \mbox{$\Omega$}%
                 \fi%
                 }%

\newcommand{\mevc}{\ifmmode \rm{MeV/c}%
                 \else%
                            \mbox{MeV/c}%
                 \fi%
                 }%

\newcommand{\mevcc}{\ifmmode \rm{MeV/c}^2%
                 \else%
                           \mbox{MeV/c$^2$}%
                 \fi%
                 }%
\newcommand{\gev}{\ifmmode \rm{GeV}%
                 \else%
                           \mbox{GeV}%
                 \fi%
                 }%

\newcommand{\gevc}{\ifmmode \rm{GeV/c}%
                 \else%
                           \mbox{GeV/c}%
                 \fi%
                 }%

\newcommand{\gevcc}{\ifmmode \rm{GeV/c}^2%
                 \else%
                           \mbox{GeV/c}$^2$%
                 \fi%
                 }%
\newcommand{\kev}{\ifmmode \rm{keV}%
                 \else%
                           \mbox{keV}%
                 \fi%
                 }%

\newcommand{\mmsq}{\ifmmode \rm{mm}^2%
                 \else%
                           \mbox{mm}$^2$%
                 \fi%
                 }%

\newcommand{\mm}{\ifmmode \rm{mm}%
                 \else%
                           \mbox{mm}%
                 \fi%
                 }%

\newcommand{\cm}{\ifmmode \rm{cm}%
                 \else%
                           \mbox{cm}%
                 \fi%
                 }%
\newcommand{\mkm}{\ifmmode \mu\rm{m}%
                 \else%
                           \mbox{$\mu$m}%
                 \fi%
                 }%

\newcommand{\m}{\ifmmode \rm{m}%
                 \else%
                           m%
                 \fi%
                 }%

\newcommand{\tl}{\ifmmode \rm{T}%
                 \else%
                           \mbox{T}%
                 \fi%
                 }%

\newcommand{\dedx}{\ifmmode \rm{dE/dx}%
                 \else%
                           \mbox{dE/dx}%
                 \fi%
                 }%

\newcommand{\tof}{\ifmmode \rm{ToF}%
                 \else%
                           \mbox{ToF}%
                 \fi%
                 }%

\newcommand{\lumi}{\ifmmode \rm{\cal{L}}%
                 \else%
                           \cal{L}%
                 \fi%
                 }%

\newcommand{\ee}{\ifmmode \rm{e}^+\rm{e}^-%
                 \else%
                          \mbox{$\rm{e}^+\rm{e}^-$}%
                 \fi%
                }%

\newcommand{\AmS}{{\protect\the\textfont2
  A\kern-.1667em\lower.5ex\hbox{M}\kern-.125emS}}

\title{Pentaquark Searches at CDF\thanks{Presented at 6th International Conference on Hyperons, Charm and Beauty Hadrons, BEACH 2004, June 27 - July 3, 2004, IIT, Chicago, U.S.A.}}

\author{Dmitry O.~Litvintsev \address[fermilab]{Fermi National Accelerator Laboratory,
	M.S. 318, P.O. Box 500, Batavia, IL, 60510, U.S.A. } (for CDF Collaboration)}

\runtitle{Pentaquark Searches at CDF}
\runauthor{D. Litvintsev}

\begin{document}

\begin{abstract}

Recently there has been revival of interest in exotic baryon spectroscopy 
triggered by experimental evidence for pentaquarks containing $u,d,s$ and $c$-quarks.
We report results of the searches for pentaquark states in decays to ${\rm pK^0_S}$, 
${\rm \Xi^-\pi^\pm}$ and ${\rm D^{*-}p}$ performed at CDF detector using 
${\rm 220~pb^{-1}}$ sample of ${\rm p\bar{p}}$ interactions at $\sqrt{s}$ of 1.96~TeV. 
No evidence for narrow resonances were found in either mode.

\vspace{1pc}
\end{abstract}

\maketitle

\section{Introduction}


Searches for states, characterized by so-called exotic quantum numbers, i.e., 
quantum numbers that cannot be obtained from minimal $(q\bar{q})$ or $(qqq)$ 
configurations of standard mesons and baryons, have taken place since the introduction 
of QCD and quark model. Until recently, these searches yielded very little conclusive 
evidence for such states. 

During 2003, however, the situation has changed dramatically.
An observation of a narrow resonance at $(1540\pm 10)~\mevcc$, called $\Theta^+$, decaying 
to ${\rm nK^+}$, produced in ${\rm \gamma n \ra K^-\Theta^+}$, was reported by LEPS~\cite{Nakano:2003qx}. 
The state has exotic quantum number, positive strangeness, and cannot exist in
a simple three-quark model. A pentaquark interpretation was employed  suggesting the quark content to be ($uudd\bar{s}$). 
This observation  was confirmed by various 
 experiments using incident beams of real and quasi-real photons, kaons and neutrinos~\cite{Barmin:2003vv}. 

Afterwords, 
followed two other manifestly exotic S=-2 baryon states decaying to 
$\Xi^-\pi^{\pm}$ 
in pp collisions of CERN SPS at $\sqrt{s}=17.2~\gev$, reported by NA49~\cite{Alt:2003vb}.  The combined mass of $\Xi_{3/2}^{--}$ and $\Xi_{3/2}^0$ was measured to be $(1.862 \pm 0.002)~\gevcc$.
These resonances were interpreted as ${\rm I_3=-3/2}$ and ${\rm I_3=+1/2}$ 
partners of isospin quadruplet of five-quark states with quark contents ($dsds\bar{u}$) and ($dsus\bar{d}$) respectively.

The ${\rm \Theta^+}$ and ${\rm \Xi^{--,0}}$ states are consistent with being members of minimal SU(3) anti-decuplet of pentaquark states 
predicted in the chiral soliton model of baryons \cite{diakonov}. 

In March 2004, the H1 experiment at HERA reported a narrow resonance, the ${\rm \Theta_c}$, 
 decaying to ${\rm D^{*-} p}$ at a mass  3.099~\gevcc~\cite{h1}.  
It can be seen as the charmed analog 
of ${\rm \Theta^+}$ with quark content ($uudd\bar{c}$).

For strongly decaying particles all reported states are unusually narrow with widths consistent 
with detector resolution. The reported masses  of ${\rm \Theta^+}$  vary widely over a  range,  between  1526 and 1542~\mevcc, for different experiments. 

All pentaquark observations are based on relatively small data samples. 
Typically  20 -- 100 signal events are seen with a 
significance in the range from 4 to 6$\sigma$. While the $\Theta^+$ has been seen by several experiments, an independent 
confirmation of the $\Xi_{3/2}^{--,0}$ and ${\rm \Theta_c}$ states has not been made. Definitely high statistics 
experiments are needed to establish pentaquark states unambiguously. 

At this conference we report results of the searches for exotic baryons ${\rm \Theta^+}$, ${\Xi^{--,0}_{3/2}}$ and 
${\rm \Theta_c}$, produced in quark fragmentation 
in ${\rm p\bar{p}}$ collisions at $\sqrt{s}=1.96$~TeV, 
using  data recorded with upgraded CDF~II detector at Tevatron, Fermilab. 

\begin{figure*}[htb]\centering
\includegraphics[width=50mm,height=35mm]{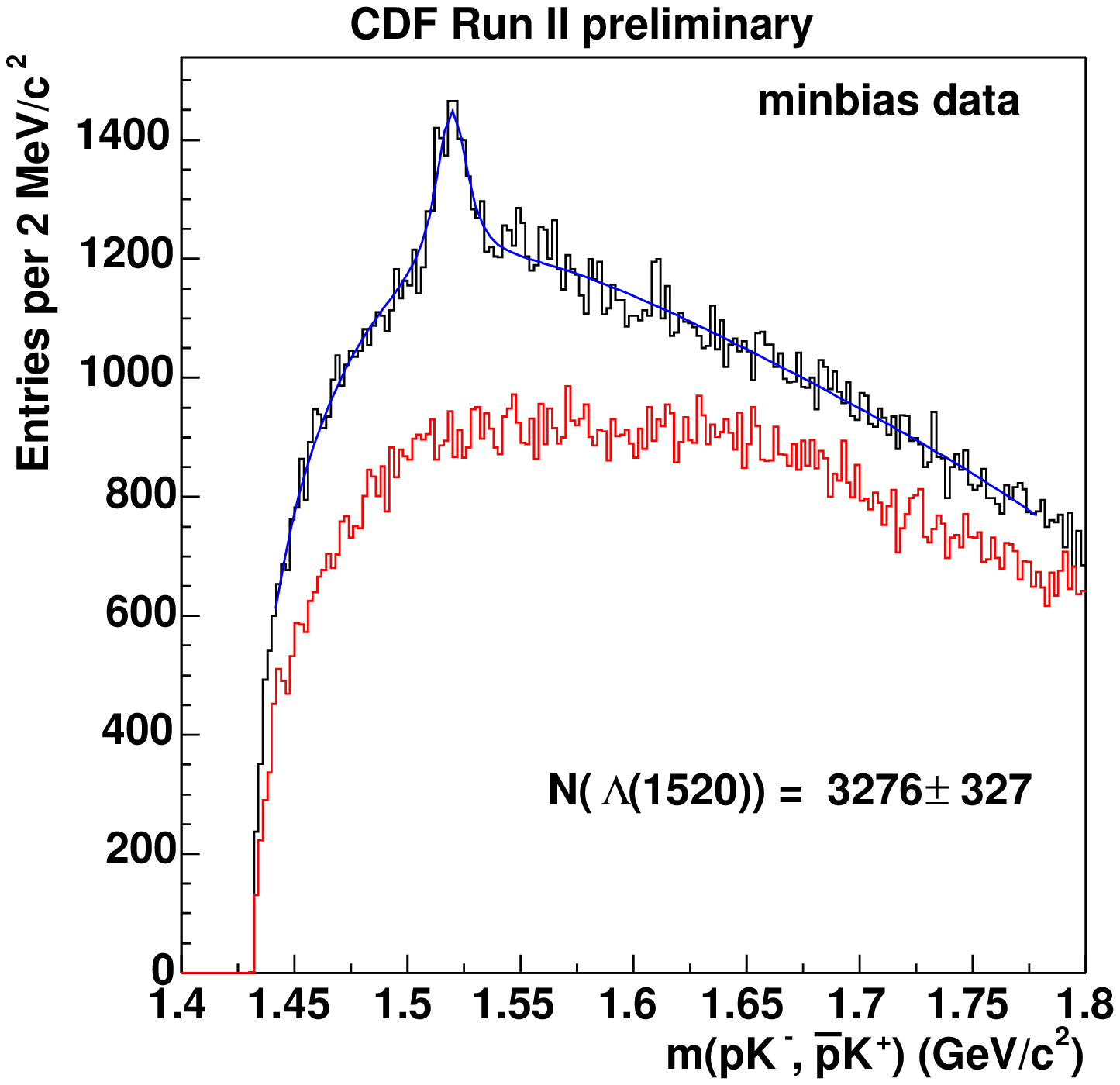}
\includegraphics[width=50mm,height=35mm]{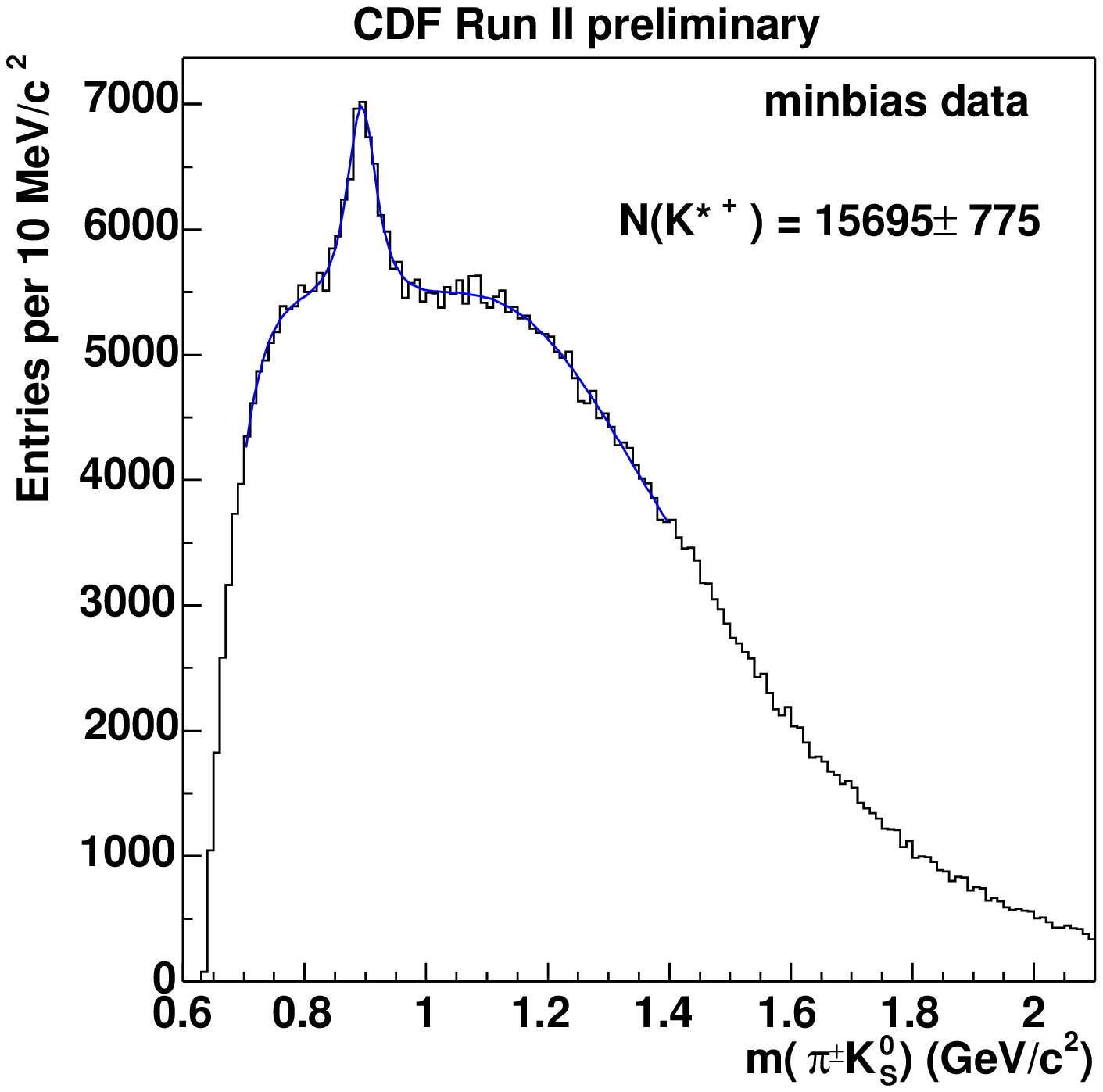}
\includegraphics[width=50mm,height=35mm]{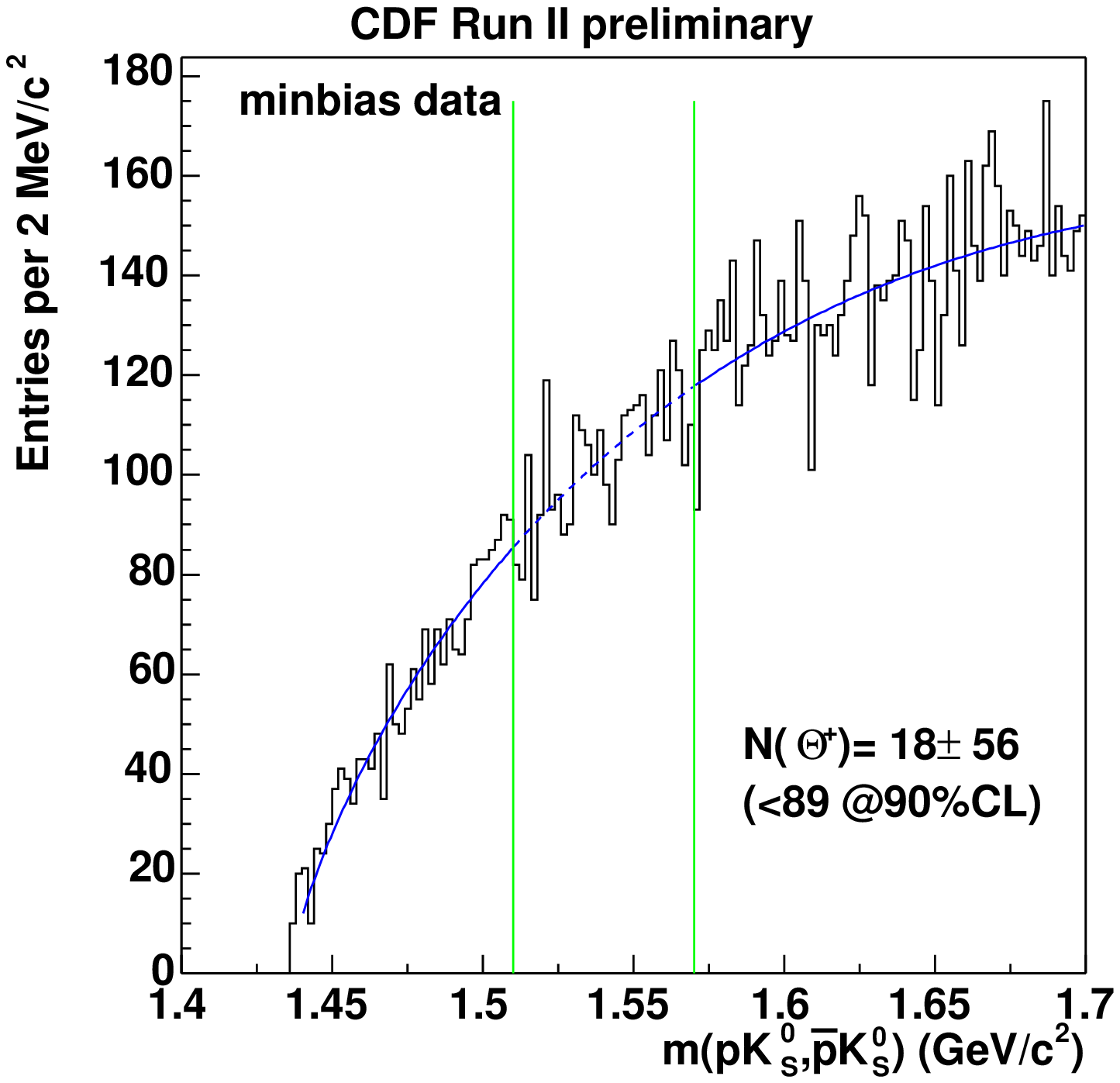}
\vspace*{-1cm}
\caption{Left: an invariant mass spectrum of the ${\rm pK^-}$ (and  ${\rm pK^+}$) combinations showing well 
	established resonance ${\Lambda(1520)}$. 
        Center: a ${\rm K^0_S\pi^+}$  invariant mass spectrum showing clear ${\rm K^{*+}}$ signal.
        Right: an invariant mass spectrum of 
	${\rm pK^0_S}$ combinations, two vertical lines indicate the ${\rm \Theta^+}$ search
	window}
\label{fig:theta}
\end{figure*}

\section{The data samples}

The data used in this analysis were obtained through three different trigger
paths.  The first data set was obtained by a trigger that was specialized for
recording hadronic B-decays, the so-called B-hadronic trigger. It requires 
the presence of two displaced oppositely charged tracks with hits in silicon,
each having  ${\rm p_T>2~\gevc}$, the scalar sum of the two transverse momenta exceeding
5.5~\gevcc\  and an impact parameter  greater than 0.1 mm. The other two complimentary
datasets were produced by the Jet20 trigger, which requires the presence of a jet with an ${\rm E_T>20~\gev}$ 
and a minimum bias trigger, which selects soft inelastic collisions. 
The data corresponded to an integrated luminosity of 220-240~pb$^{-1}$. The Jet20 and the minimum bias triggers 
are heavily prescaled resulting in effective integrated luminosities ${\rm 0.36~pb^{-1}}$ and ${\rm 0.37~nb^{-1}}$
correspondingly. 

\section{Search for ${\rm \Theta^+\ra pK^0_S}$}

In contrast to ${\rm nK^+}$ channel, the decay ${\rm \Theta^+\ra pK^0}$ does not provide manifestly exotic signature 
as the ${\rm K^0}$ is reconstructed as ${\rm K^0_S}$.  However observation or non-observation 
of a narrow state decaying to  ${\rm pK^0_S}$ at the same mass as in ${\rm nK^+}$ channel 
can be interpreted as positive confirmation or evidence against the existence of ${\rm \Theta^+}$ as 
this state is expected to decay to  ${\rm \Theta^+\ra pK^0}$ and  ${\rm nK^+}$ with the same rate. 

	\begin{table}[bth]\centering
	\vspace*{-0.5cm}
		\caption{Event yields of reconstructed resonances and 90\% CL limits on $\Theta^+$ yields} 		\begin{tabular}{c|c|c}\hline
        	Resonance & Minbias data & Jet20 data \\        \hline
		${\rm \Lambda(1520)}$       & $3276 \pm 327$ & $4915\pm702$ \\ 
	        ${\rm K^{*+}}$              & $15695\pm 775$ & $35769 \pm 1390$ \\ 
		${\rm \Theta^+}$            & $18\pm56$       & $-56\pm100$ \\ 
		90\% CL on ${\rm \Theta^+}$ & $<89$           & $<76$ \\ \hline
		\end{tabular}
		\label{table:theta-ul}
	\vspace*{-0.8cm}
	\end{table}

\begin{figure*}[ht]\centering
\includegraphics[width=52mm,height=40mm]{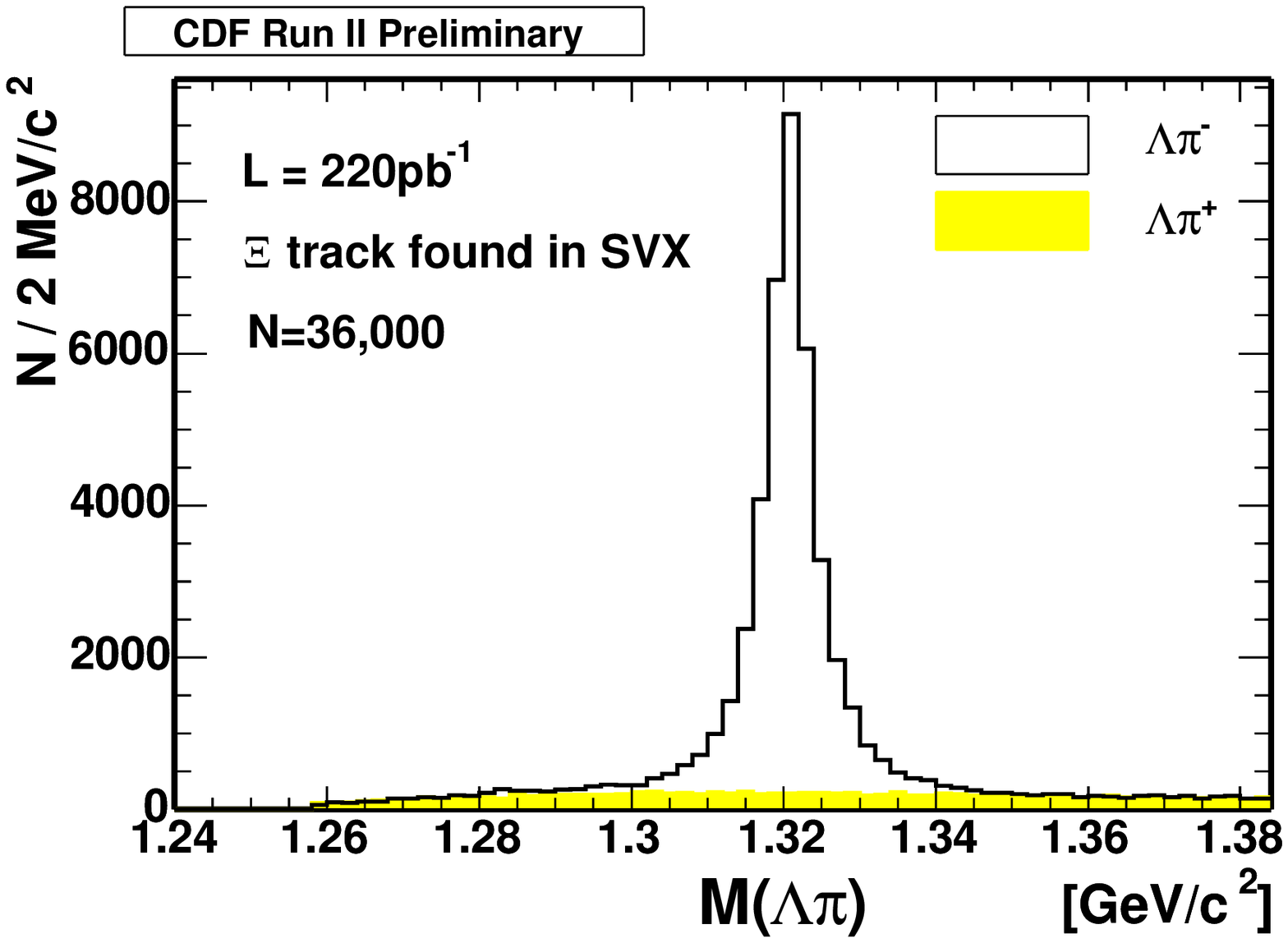}
\includegraphics[width=52mm,height=40mm]{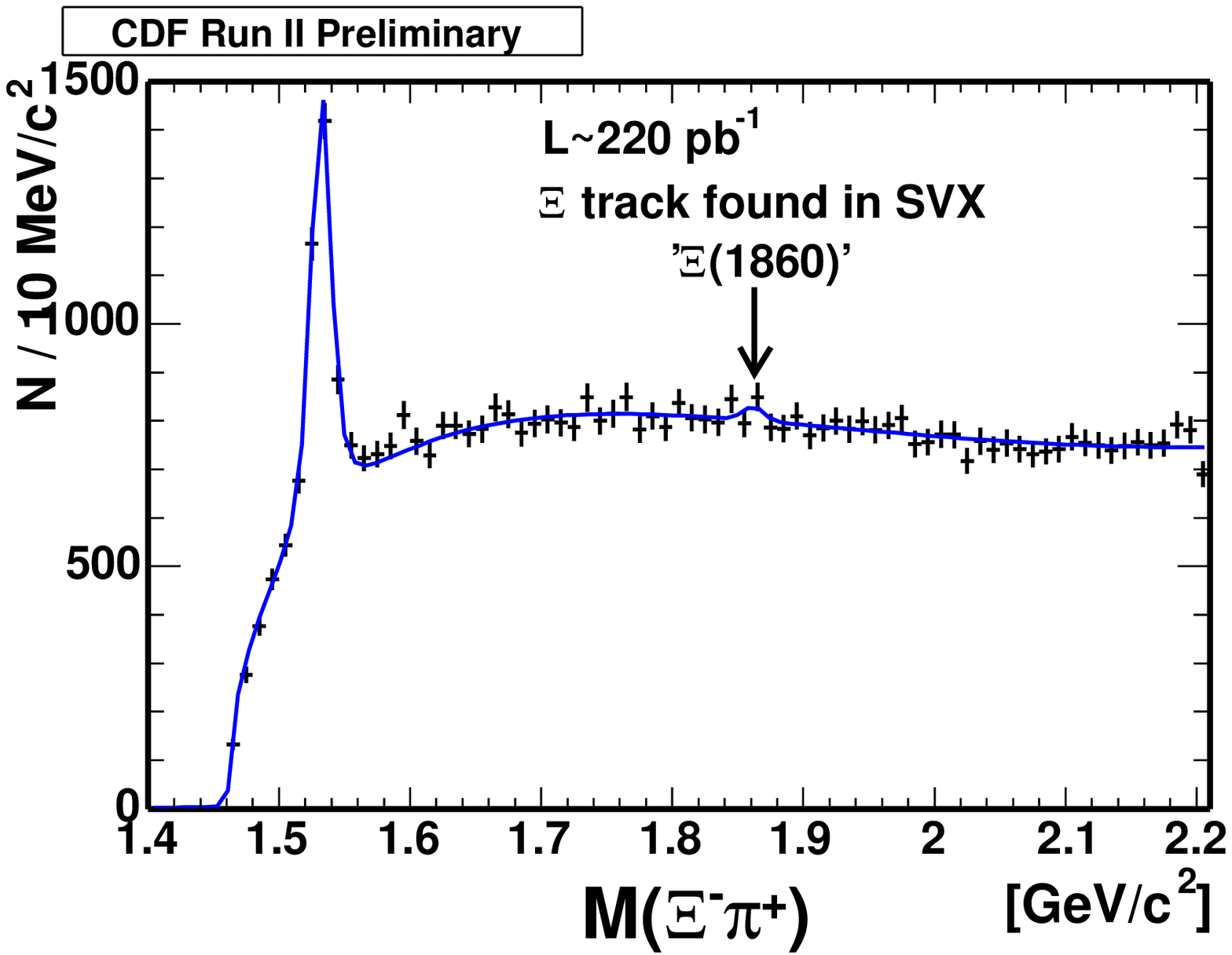}\includegraphics[width=52mm,height=40mm]{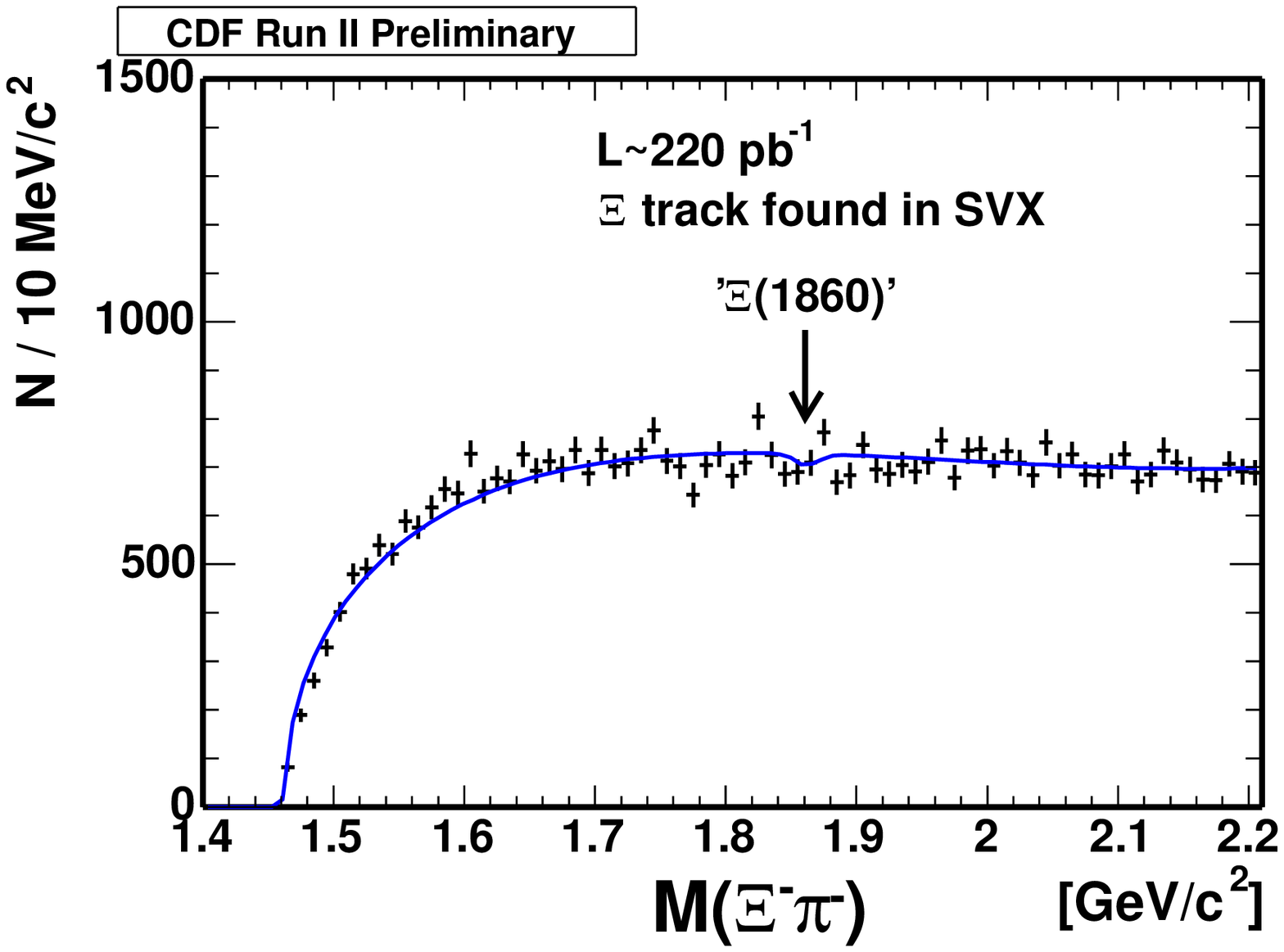}
\vspace*{-1cm}
\caption{Left: an invariant mass spectrum of ${\rm \Lambda\pi}$ combinations that have associated 
	       $\Xi$ track in SVX, open histogram shows right sign $\Lambda\pi^-$ combinations, 
		filled histogram shows wrong sign,   $\Lambda \pi^+$  combinations.
	 Center: an invariant mass of $\Xi^-\pi^+$ combinations. Right:  an invariant mass of $\Xi^-\pi^-$ combinations. Arrows 
	mark the mass at 1862~\gevcc.}
\label{fig:ttt-xi}
\end{figure*}

	\begin{table*}[bth]\centering
			\caption{Event yields of ${\rm \Xi(1530)}$  
                                 and upper limits at 90\%CL on event yields and relative rates $\Xi^{--}_{3/2}$ and $\Xi^{0}_{3/2}$
                                 assuming equal detector acceptance of  ${\rm \Xi^{--,0}_{3/2}}$ and ${\rm \Xi^(1530)}$.}
			\label{table:xi}
		\begin{tabular}{c|c|c}\hline
		                          & Hadronic trigger               & Jet20 data \\ \hline 
		${\rm N(\Xi)}$            & $35722\pm326$                  & $4870\pm122$  \\ 
		${\rm N(\Xi(1530))}$      & $2182\pm92$                    & $387\pm34$ \\ 
		${\rm N(\Xi^{--}_{3/2})}$ & ${\rm -54\pm47~(<51~@~90\%~CL)}$ & ${\rm -4\pm18~(28~@~90\%~CL)}$ \\   
		${\rm N(\Xi^{0}_{3/2})}$  & ${\rm 57\pm51~(<126~@~90\%~CL)}$ & ${\rm -14\pm19~(25~@~90\%~CL)}$ \\    
		$\frac{{\rm \sigma(\Xi^{--}_{3/2})\cdot Br(\Xi\pi^-)}}
		{{\rm \sigma(\Xi(1530))\cdot  Br(\Xi\pi^+)}}$ & ${\rm <0.03~@~90\%~CL}$           & ${\rm <0.07~@~90\%~CL}$ \\ 
		$\frac{{\rm \sigma(\Xi^{0}_{3/2})\cdot Br(\Xi\pi^+)}}
		{{\rm \sigma(\Xi(1530))\cdot  Br(\Xi\pi^+) }} $ & ${\rm <0.06~@~90\%~CL}$ & ${\rm <0.06~@~90\%~CL}$ \\ \hline
		\end{tabular}
	\vspace*{-0.5cm}
	\end{table*}

The information from TOF system has been used to produce clean samples of protons. Protons were identified as 
tracks having ${\rm 2\sigma}$ separation between TOF assuming proton and kaon hypotheses. 
Figure~\ref{fig:theta} shows from left to right invariant mass spectra of ${\rm pK^-}$, ${\rm K^0_S\pi^+}$ and ${\rm pK^0_S}$
combinations in the minimum bias dataset. Clear signals from known resonances ${\Lambda(1520)}$ and ${\rm K^{*+}}$ can be seen. There is no indication 
of a narrow state in ${\rm pK^0_S}$ mass spectrum. 
The spectrum was 
fit to the  background function with the search mass window ${\rm 1.510<M(pK^0_S)<1.570~\gevcc}$ excluded from the fit. 
Table~\ref{table:theta-ul} summarizes the yields of known resonances and 90\% CL limits on $\Theta^+$ yields in the minimum bias  data and in the Jet20 data.

\section{Search for ${\rm \Xi^{--,0}_{3/2}}$}

\begin{figure*}[ht]\centering
\includegraphics[width=39mm,height=40mm]{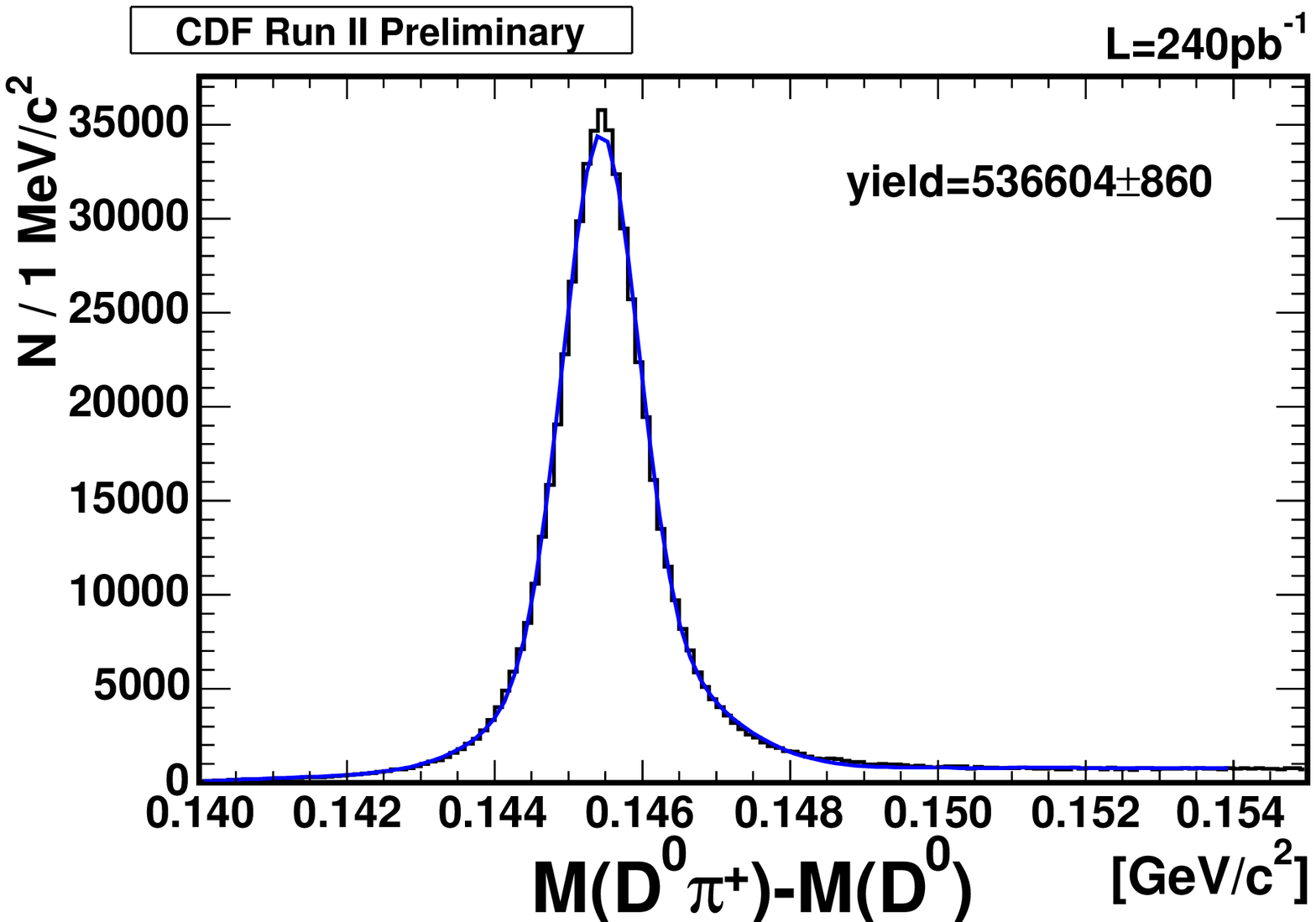}
\includegraphics[width=39mm,height=40mm]{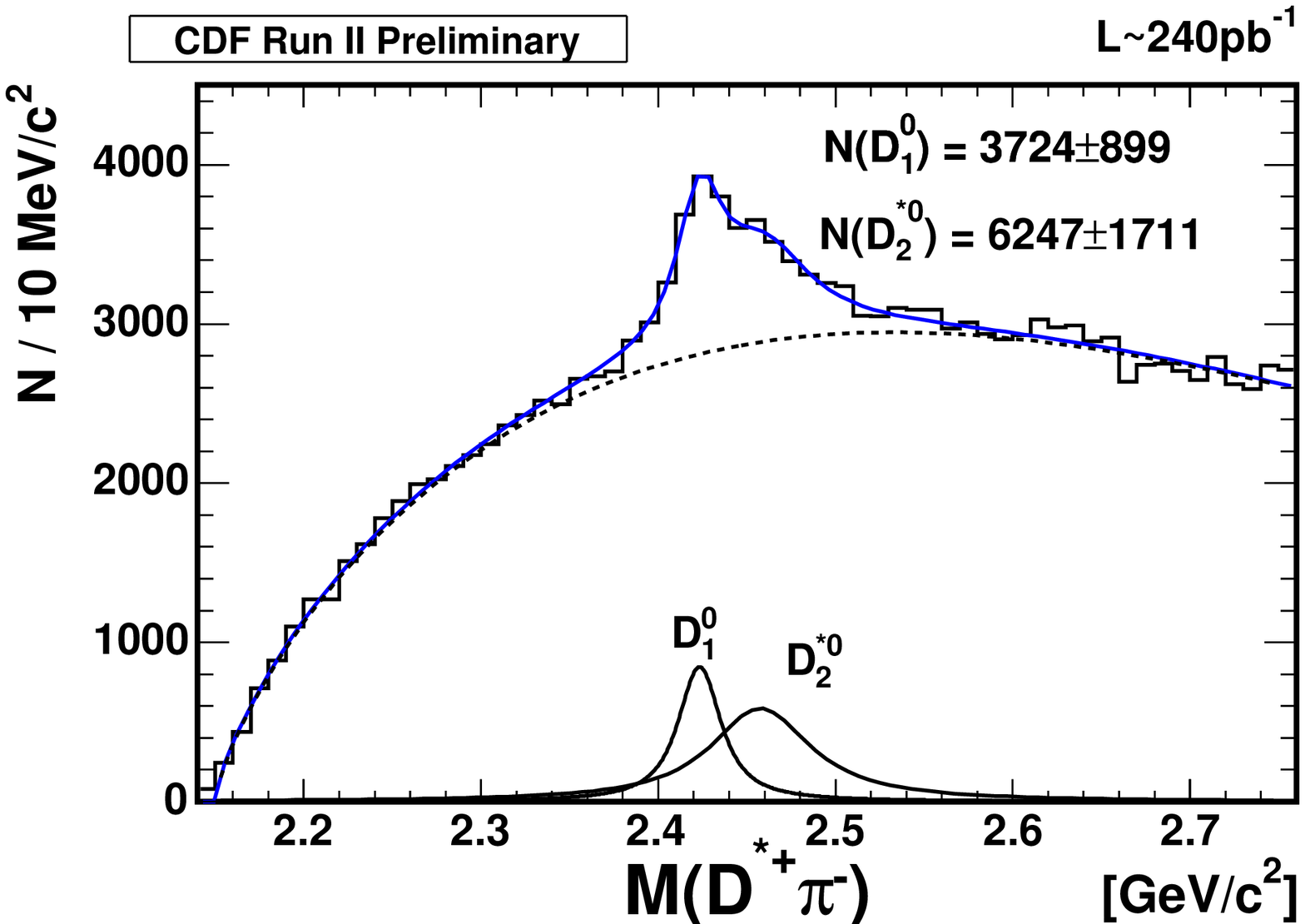}
\includegraphics[width=39mm,height=40mm]{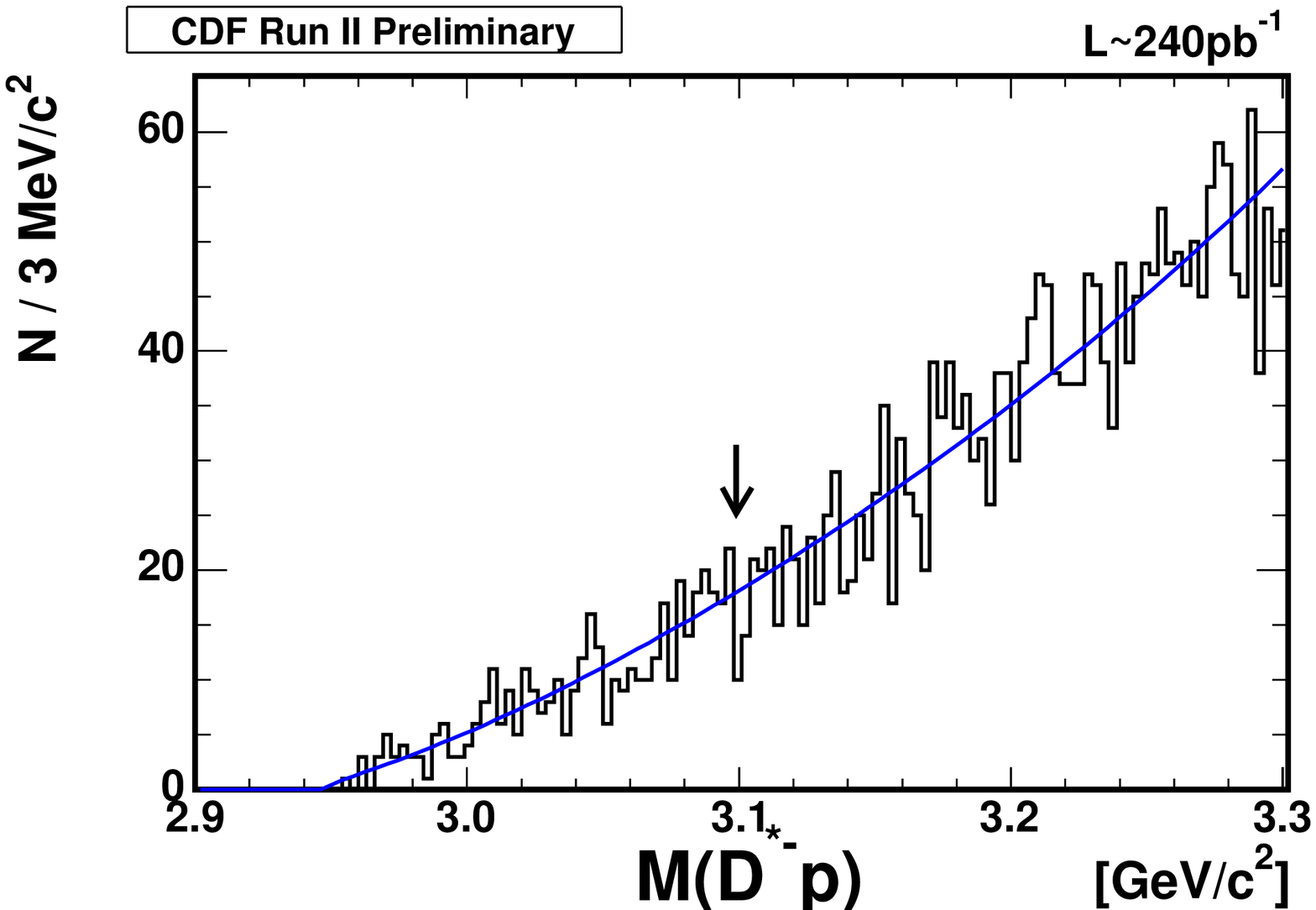}
\includegraphics[width=39mm,height=40mm]{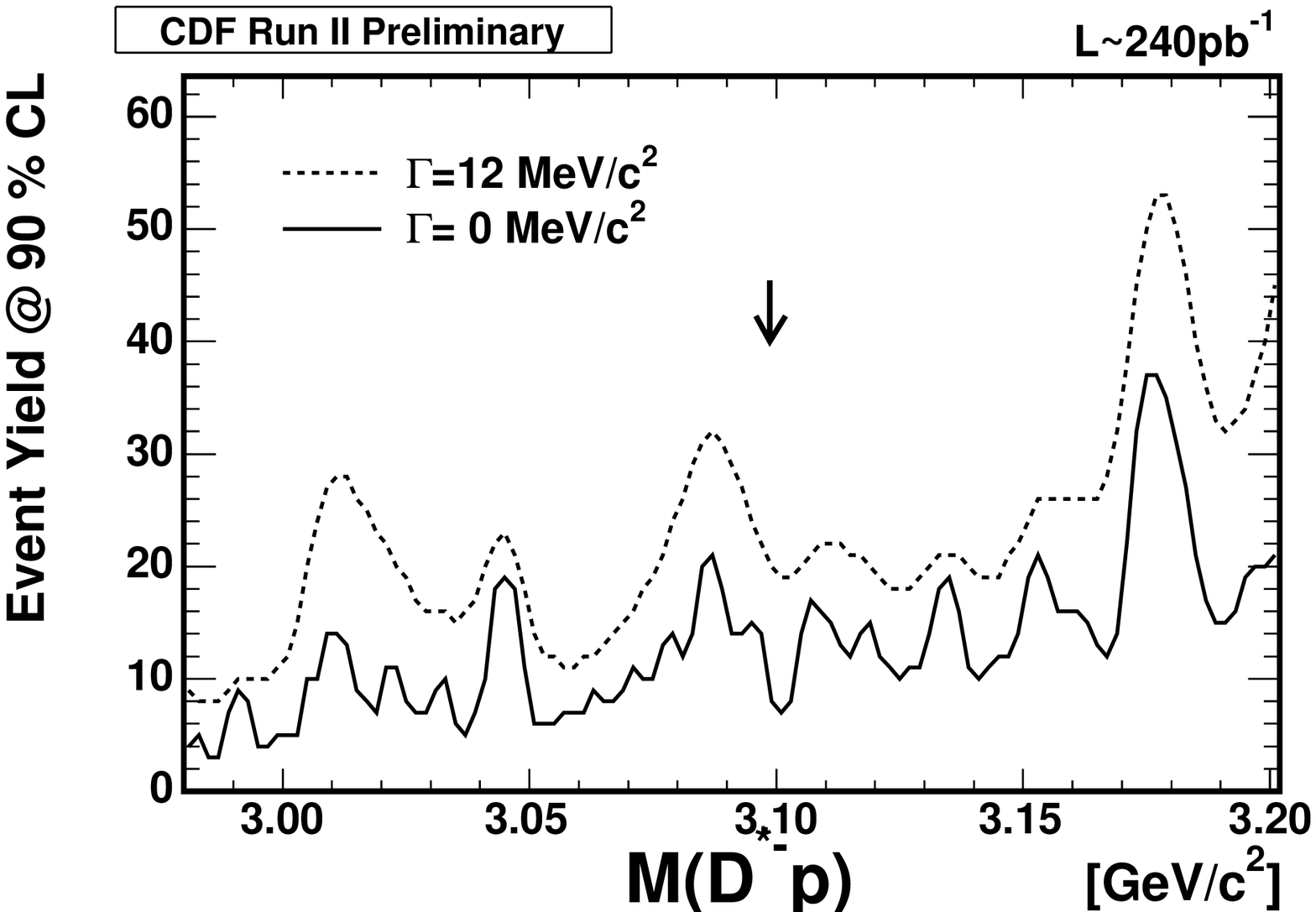}
\rput(-6.8,3){a)}
\rput(-2.8,3){b)}
\rput(1.5,3){c)}
\rput(5.5,3){d)}
\vspace*{-1cm}
\caption{a) ${\rm D^0\pi^+}$ mass difference spectrum,
         b) ${\rm \Delta M(D^{*+} \pi^-)+2.01}$ spectrum,
	 c) ${\rm \Delta M(D^{*-}  p)+2.01}$ mass spectrum,
	 d) upper limits on event yield vs mass obtained assuming zero natural width of 
	    ${\rm \Theta_c}$ (solid line) and 12~\mevcc\ (dashed line). Arrow marks the mass at 3.099~\gevc.}
\label{fig:dst}
\end{figure*}

The pentaquark candidates were reconstructed through the decay chain:
$\Xi_{3/2}^{--,0} \ra  \Xi^{-}~\pi^{-,+}$, $\Xi^{-} \ra \Lambda \pi^{-}$, 
$\Lambda \ra p \pi^{-}$.
The $\Lambda$ candidates were reconstructed from 
oppositely charged pairs of tracks. The track with the highest transverse momentum in the pair was assigned the proton mass.

The long lifetime of the $\Xi^{-}$ and the ${\rm p_T}$ requirements 
produce a  decay point of the $\Xi^{-}$ 
that lies outside the SVX~II volume in a significant
fraction of  events.
This  makes it possible to reconstruct the $\Xi^{-}$ track from the 
hits that  the particle has left traversing the layers of SVX~II.
The position of the $\Xi^{-}$ decay and its momentum information obtained
in the mass constrained fit 
were used to define a road for a special reconstruction algorithm that tracks the $\Xi^{-}$ in
the silicon system.
This algorithm required a minimum of two axial SVX hits on the track 
within the radius defined by the measured decay point.  
A plot on the left of the  Figure~\ref{fig:ttt-xi}  shows
the $\Lambda\pi^-$ invariant mass of $\Xi^{-}$ candidates that have 
associated SVX tracks.  A very clean $\Xi^{-}$ signal containing 36,000 events is visible. 
The $\Xi^{-}$ sample used in subsequent analysis was limited to candidates that
were successfully tracked in the silicon detector, and had  invariant mass within 
10~\mevcc\ of the world average $\Xi^{-}$ mass.

To search for new states, the $\Xi^{-}$ tracks in each event were combined with the remaining 
tracks.  These tracks were assigned the pion mass, and were refit with the
constraints that they form a good vertex with \Cascade\ track  and the total 
momentum of the pair was constrained to point back to the primary vertex.

The invariant mass spectra of $\Xi^{-}\pi^+$ and $\Xi^{-}\pi^-$ are  shown 
in Figure~\ref{fig:ttt-xi}.  A prominent peak corresponding to the decay 
$\Xi(1530)\ra\Xi^{-}\pi^+$ can be clearly seen.  No other peaks are visible in 
either $\Xi^{-}\pi^+$ nor $\Xi^{-}\pi^-$ spectra. The expected detector resolution 
at 1862~\gevcc\  is 8~\mevcc. Table~\ref{table:xi} summarizes the measurements for hadronic sample 
and Jet20 sample.

\section{Search for ${\rm \Theta_c^0 \ra D^{*-} p}$}

This mode is especially attractive as two tracks from displaced ${\rm D^0}$ vertex come in 
on unique CDF B-hadronic trigger. In the H1 pentaquark paper they mentioned that the fraction
of ${\rm \Theta_c^0}$ is roughly 1\% of the total ${\rm D^*}$ production \cite{h1}. Assuming that 
the dominant production mechanism of ${\rm \Theta_c^0}$ in deep inelastic 
${\rm ep}$ collisions is fragmentation of the c-quark produced in ${\rm \gamma^*p\ra c\bar{c}}$, 
the fragmentation probability $f({\rm \bar{c}\ra\Theta_c^0})$ could be  $2.35\times 10^{-3}$. 
If this was true then CDF should see $\sim 10^4$ per ${\rm 100~pb^{-1}}$~\cite{frag}.

	\begin{table}[bth]\centering
			\caption{Event yields  and upper limits at 90\%CL on event yields  of ${\rm \Theta_c}$ }
			\label{table:dst}
                \begin{tabular}{|l|l|}\hline
	Channel & yield  \\ \hline
	${\rm N(D_2^{*0})\ra D^{*+}\pi^- }$  &  $6247 \pm 1711$  \\
	${\rm N(D_1^{0})\ra D^{*+}\pi^- }$   &  $3724 \pm 899$  \\
	${\rm \Theta_c\ra D^{*-} p}$         &  ${\rm <21~@~90~\%~CL}$ \\ \hline
\end{tabular}
\vspace*{-1.0cm}
\end{table}

The CDF experiment has accumulated 540,000 ${\rm D^{*+}}$ candidates, very cleanly
 reconstructed in ${\rm D^0\pi^+}$, ${\rm D^0\ra K^-\pi^+}$ chain (see Figure~\ref{fig:dst}(a)). An ability 
to reconstruct reference ${\rm D^{**}\ra D^{*+}\pi^-}$ channel is demonstrated in Figure~\ref{fig:dst}(b). 

The proton identification procedure was based on combined likelihood ratio calculated from measurements of 
specific ionization and time-of-flight for allowed mass hypotheses ($e,\mu,\pi$,K and p). A track was treated 
as proton if the corresponding likelihood ratio exceeded 40\%. 

The ${\rm D^{*-} p}$ mass difference spectrum ${\rm M(D^{*-}p) = \Delta M(D^{*-}p)+2.01}$ is presented 
in Figure~\ref{fig:dst}(c), there is no apparent narrow resonance signal. 
The expected mass resolution at 3.099~\gevcc\  is $2.3~\mevcc$. The function superimposed in 
Figure~\ref{fig:dst}(c) shows the result of the unbinned log-likelihood fit with the background function 
consisting of square root threshold function multiplied by the 3rd order polynomial. A series of such fits were 
performed with ${\rm \Theta_c^0}$ signal introduced as Breit-Wigner resonance function convoluted with a  Gaussian resolution 
function. The mass of the state was varied from 2.98 to 3.2~\gevcc\  with 2~\mevcc\  step.  The resulting functions of 
90\% CL event yield limits vs ${\rm D^{*-} p}$ mass are shown in Figure~\ref{fig:dst}(d) for two cases -- assuming 
zero natural width  and assuming $12~\mevcc$ natural width of the resonance. As the final result we quote 
the worst upper limit obtained within the allowed search window $(3.099 \pm 0.018)~\mevcc$. All results are summarized in 
Table~\ref{table:dst}.

\section{Conclusion}

Using high  statistics samples of ${\rm K^0_S}$, $\Xi^-$ and ${\rm D^{*+}}$ produced in 
${\rm p\bar{p}}$ interactions at $\sqrt{s}$=1.96~TeV and reconstructed with the detector having 
excellent momentum resolution and a powerful particle identification system 
we performed a sensitive search for narrow pentaquark states
decaying to  ${\rm pK^0_S}$, ${\rm \Xi^-\pi^\pm}$ and ${\rm D^{*-}p}$. No signal found in either channel. 
The analysis technique was tested by demonstrating the ability to reconstruct signals from standard narrow resonances 
having similar decay kinematics, $\Lambda(1520)$, ${\rm K^{*+}}$, $\Xi(1530)$, ${\rm D^{**}}$. 
The absence of pentaquark signals means that, if existent, pentaquarks 
are not produced in the process of quark fragmentation or such production is severely suppressed.


\begin{thebibliography}{10000}

\begin{small}

\bibitem{Nakano:2003qx} T.~Nakano {\it et al.}, Phys.\ Rev.\ Lett.\  {\bf 91}, 012002 
\bibitem{Barmin:2003vv} V.~V.~Barmin {\it et al.}, Phys.\ Atom.\ Nucl.\  {\bf 66}, 1715; 
	J.~Barth {\it et al.}   Phys.\ Lett.\ B {\bf 572}  127; 
         A.~E.~Asratyan {\it et al.}, Phys.\ Atom.\ Nucl.\  {\bf 67}, 682; 
         V.~Kubarovsky {\it et al.}, Phys.\ Rev.\ Lett.\  {\bf 92}, 032001;
         A.~Airapetian {\it et al.}, Phys.\ Lett.\ B {\bf 587} 213 
\bibitem{Alt:2003vb} C.~Alt {\it et al.},   Phys.\ Rev.\ Lett.\  {\bf 92}, 042003
\bibitem{diakonov} A.~Diakonov {\it et al.},  Z.\ Phys.\  A {\bf 359 } 305 
\bibitem{h1} A.~Aktas {\it et al.}, hep-ex/0403017
\bibitem{frag} K.~Cheung,  arXiv:hep-ph/0405281

\end{small}
\end{thebibliography}
\end{document}